\documentclass[11pt,letterpaper]{article}
\usepackage[utf8]{inputenc}
\usepackage{ dsfont }

\pdfoutput=1
\usepackage{color}
\definecolor{darkblue}{rgb}{0.1,0.1,.7}
\usepackage[colorlinks, linkcolor=darkblue, citecolor=darkblue, urlcolor=darkblue, linktocpage]{hyperref}
\usepackage[]{amsmath}
\usepackage[]{graphicx}
\usepackage[]{latexsym}
\usepackage[utf8]{inputenc}
\usepackage{slashed,graphicx,color,amsmath,amssymb}
\usepackage[mathscr]{eucal}
\usepackage{mathrsfs}
\usepackage[margin=10pt,font=small,labelfont=bf]{caption}
\usepackage{amscd}
\usepackage{bm}
\usepackage{xcolor}
\usepackage{upgreek}
\usepackage[square, comma, sort&compress,numbers]{natbib}
\usepackage[all,cmtip]{xy}
\usepackage[margin=1.7in]{geometry}
\usepackage{cleveref}
\usepackage[color=cyan!30!white,linecolor=red,textsize=footnotesize]{todonotes}
\usepackage{soul}
\numberwithin{equation}{section}
\definecolor{shadecolor}{rgb}{0.9,0.9,0.95}

\hypersetup{
	pdfstartview={FitH},    
	pdftitle={Adding Flavor to the Narain ensemble},    
	pdfauthor={Datta, Duary, Kraus, Maity, Maloney},     
	colorlinks=true,       
	linkcolor=blue,          
	citecolor=blue!59!black! ,        
	filecolor=magenta,      
	urlcolor=blue           
}

\usepackage[utf8]{inputenc}
\usepackage{tikz}
\usetikzlibrary{shapes.misc}
\usetikzlibrary{decorations.markings}
\tikzset{cross/.style={cross out, draw=black, ultra thick, minimum size=2*(#1-\pgflinewidth), inner sep=0pt, outer sep=0pt},
cross/.default={5pt}}
\usetikzlibrary{decorations.pathmorphing}

\tikzset{snake it/.style={decorate, decoration=snake}}

\usetikzlibrary{arrows}
\usetikzlibrary{decorations.markings}
\tikzset{
  big arrow/.style={
    decoration={markings,mark=at position 1 with {\arrow[scale=2.5,#1]{>}}},
    postaction={decorate},
    shorten >=0.4pt},
  big arrow/.default=blue}
  \tikzset{
  double arrow/.style={
    decoration={markings,mark=at position 1 with {\arrow[scale=2.5,#1]{>>}}},
    postaction={decorate},
    shorten >=0.4pt},
  big arrow/.default=blue}
  
\usetikzlibrary{calc} 

\def\be{\begin{equation}}
\def\ee{\end{equation}}

\DeclareFontFamily{U}{wncy}{}
\DeclareFontShape{U}{wncy}{m}{n}{<->wncyr10}{}
\DeclareSymbolFont{mcy}{U}{wncy}{m}{n}
\DeclareMathSymbol{\Sha}{\mathord}{mcy}{"58}

 \makeatletter
 \g@addto@macro\bfseries{\boldmath}
 \makeatother

\makeatletter
\if@todonotes@disabled

\else

\fi
\makeatother

\begin{document}

\definecolor{tinge}{RGB}{255, 244, 195}
\sethlcolor{tinge}
\setstcolor{red}

\vspace{.2in} {\Large
\begin{center}
{\LARGE Perturbative Mellin amplitudes of $\mathcal{N}=4$ SYM} 
\end{center}}
\vspace{.2in}
\begin{center}
{Faizan Bhat$^a$ and Pronobesh Maity$^b$$^,$$^c$}
\\
\vspace{.3in}
\small{
$^a$  \textit{Centre  for  High  Energy  Physics,  Indian  Institute  of  Science, \\ C.V.  Raman  Avenue,  Bangalore  560012,  India}\\
\vspace{.2cm}
$^b$\textit{International Centre for Theoretical Sciences,\\
	Shivakote, Hesaraghatta Hobli, Bengaluru North 560 089, India}
\\
\vspace{0.2cm}
$^c$\textit{Laboratory for Theoretical Fundamental Physics, EPFL,\\
Rte de la Sorge, Lausanne, Switzerland.}
}\vspace{0cm}


\end{center}

\vspace{.5in}

\begin{abstract}
\normalsize
 We compute the Mellin amplitude of the planar four-point correlator of weight-two half-BPS operators in $\mathcal{N}=4$ SYM at one and two-loop orders in the small 't Hooft coupling expansion. The two loop Mellin amplitude has an infinite number of poles, as expected from a stringy bulk dual. We then perform a Mellin conformal block expansion of the amplitude and extract the one-loop anomalous dimensions and OPE coefficients of the exchanged twist-two singlet multiplets. Our results match those of Dolan and Osborn \cite{Dolan:2004iy} but the computation is rather straightforward in Mellin space.

\end{abstract}

\vskip 1cm \hspace{0.7cm}

\newpage

\setcounter{page}{1}

\noindent\rule{\textwidth}{.1pt}\vspace{-1.2cm}
\begingroup
\hypersetup{linkcolor=black}
\tableofcontents
\endgroup
\noindent\rule{\textwidth}{.2pt}

\section{Introduction}

Starting from the seminal work of Mack \cite{Mack:2009mi}, we have seen substantial evidence \cite{Penedones:2010ue, Paulos:2011ie, Fitzpatrick:2011ia, Costa:2012cb} that `Mellin space' is the natural `Momentum space' analogue to formulate conformal field theories. In the case of quantum field theories, working in momentum space makes the Feynman rules easily tractable, and also manifests the consequences of locality, causality and unitarity. This advantage is not present for CFTs due to the absence of massive single particle states, mass being a dimensionful parameter. In a CFT, we can perform radial quantization wherein the dilatation operator acts as the hamiltonian and has discrete eigenvalues. This discrete set of operators actually appear in the operator product expansions as the exchanged conformal primaries and descendants. The Mellin amplitude corresponding to a CFT correlator makes this discrete spectrum manifest through it's poles:
\begin{equation}
    M(s,t)\sim C_{12k} C_{34k} \, \frac{Q_{l_k,m}(t)}{s-\Delta_k+l_k-2m},\quad m=0,1,2,\cdots
\end{equation}
Here $M(s,t)$ is a Mellin ampltide for four point correlator of local operators $\langle O_1 O_2 O_3 O_4\rangle$, and the poles appear at the twists $\tau_k=(\frac{\Delta_k-l_k}{2}+m)$ of the exchanged primaries and descendants at level $m$. 

For large $N$ CFTs with weakly-coupled local bulk dual, the planar Mellin amplitudes should be particularly simple, due to the large $N$ factorization of the CFT and the presence of only a few low-dimension single-trace operators in accordance with locality \cite{Heemskerk:2009pn}. In this regime of holography, Mellin amplitudes have been shown to reproduce the scattering amplitudes in the dual bulk supergravity theory, both at planar \cite{Penedones:2010ue, Rastelli:2017udc}, and also at non-planar levels \cite{Aharony:2016dwx, Alday:2018pdi, Alday:2021ajh}. There are also attempts \cite{Bekaert:2016ezc, Ponomarev:2017qab} to utilize the Mellin amplitudes to decipher the bulk dual of free CFTs in the context of higher spin holography. Indeed in the strong version of the Maldacena conjecture \cite{Maldacena:1997re}, we do expect the Mellin amplitude of weakly-coupled holographic CFTs to capture an infinite number of single-trace exchanges in the OPEs corresponding to infinte number of massive stringy modes in the bulk. The direct way to analyse the bulk in this limit is notoriously difficult due to the presence of RR flux, where the conventional local worldsheet CFT description breaks down. See \cite{Gaberdiel:2021qbb} for a recent proposal of a dual worldsheet description of free $\mathcal{N}=4$ SYM.

In this note, working in the small 't Hooft coupling and planar limit of $\mathcal{N}=4$ SYM, we compute the four-point Mellin amplitude of weight-two half-BPS operators (also known as $20^{\prime}$ operators) at one and two-loop orders.  We observe that the two-loop Mellin amplitude has an infinite number of poles, as expected from a stringy bulk dual. We then move on to expand this amplitude in Mellin conformal blocks, and extract the anomalous dimensions and OPE coefficients of the exchanged twist-two singlet multiplets at one loop. Our results match precisely with what was found in \cite{Dolan:2004iy}, but in a straightforward way using orthogonality relations of Hahn polynomials. The easiness of the analysis might be indicative of naturalness of Mellin space to formulate CFTs in general, as we discussed above.

Our motivation for this work was to understand how to extract the stringy bulk physics from their weakly-coupled dual CFTs, where we don't have an independent bulk description due to complications induced by RR flux mentioned above. We hope this note would be helpful for future developments to find a more direct way to analyse the dual string theory from perturbative large N gauge theories, perhaps connecting with the other efforts made in \cite{Gopakumar:2005fx, Gaberdiel:2020ycd, Bhat:2021dez, Fleury:2016ykk, Aprile:2020luw}.

\section{Correlator of four $20^{\prime}$ Operators}\label{half-BPS}
In $\mathcal{N}=4$ Supersymmetric Yang Mills theory, we have a tower of half-BPS operators $\mathcal{O}^{(p)}$, with $p=2,3,...$ of dimension $\Delta=p$, which transform in the $[0,p,0]$ representation of the $SU(4)$ R-symmetry subgroup of $\text{PSU}(2,2|4)$ global symmetry of the theory.. We can represent such operators as:
\begin{equation}
    \mathcal{O}^{(p)}(x,t)=t_{i_1}...t_{i_p}\,\text{tr}[\phi^{i_1}(x)...\phi^{i_p}(x)]
\end{equation}
where $t_{I}$ is six-dimensional null polarization vector which contracts the $SU(4)$ indices.  We will consider the four-point function of length two half-BPS operators, also known as {$20^{\prime}$} operators 
     \begin{equation}
        \langle  \mathcal{O}^{(2)}(x_1,t_1) \mathcal{O}^{(2)}(x_2,t_2) \mathcal{O}^{(2)}(x_3,t_3) \mathcal{O}^{(2)}(x_4,t_4)\rangle=\left(\frac{t_1.t_2t_3.t_4}{x_{12}^2x_{34}^2}\right)^2\,\mathcal{G}(u,v;\sigma,\tau)
    \end{equation}
where conformal and R-symmetry cross ratios $(u,v)$ and $(\sigma,\tau)$ are defined as
\begin{equation}
    \begin{split}
        &u=\frac{x_{12}^2\,x_{34}^2}{x_{13}^2\,x_{24}^2},\quad v=\frac{x_{14}^2\,x_{23}^2}{x_{13}^2\,x_{24}^2}\\ & \sigma=\frac{(t_1\cdot t_3)  (t_2\cdot t_4)}{(t_1\cdot t_2) (t_3 \cdot t_4)},\quad \tau =\frac{(t_1\cdot t_4) (t_2\cdot t_3)}{(t_1\cdot t_2) (t_3\cdot t_4)}
    \end{split}
\end{equation}
The fermionic generators of supergroup $\text{PSU}(2,2|4)$ imposes further constraints on the four-point function, encoded in the superconformal ward identity as:
\begin{equation}
\begin{split}\label{Superconformal Ward Identity}
   & \mathcal{G}(u,v;\sigma,\tau)=\mathcal{G}_{f}(u,v;\sigma,\tau)+S(u,v;\sigma,\tau)\,\mathcal{H}(u,v)\\ & S(u,v;\sigma,\tau)=v+\sigma^2uv+\tau^2u+\sigma v(v-1-u)+\tau (1-u-v)+\sigma \tau u(u-1-v) 
   \end{split}
\end{equation}
Here $\mathcal{G}_{f}(u,v;\sigma,\tau)$ comes from free SYM, and all the dynamical information of the correlator is encoded in $\mathcal{H}(u,v)$, which admits a conformal partial wave $G^{(l)}_{\Delta}(u,v)$ expansion as ($t'=\frac{1}{2}(\Delta-l)$):
\begin{equation}
   \mathcal{H}(u,v)=\sum_{t'}\sum_{l}A_{00t' l}u^{t'}G^{(l)}_{t'+2}(u,v)
\end{equation}

\subsection*{Position Space Results}
We can expand $\mathcal{H}(u,v)$ in 't Hooft coupling $\lambda=g_{YM}^2 N$ as
\begin{equation}
    \mathcal{H}(u,v)=\mathcal{H}^{(0)}(u,v)+\frac{\lambda}{N^2}\mathcal{H}^{(1)}(u,v)+\frac{\lambda^2}{N^2}\mathcal{H}^{(2)}(u,v)+...
\end{equation}
where the perturbative results $\{\mathcal{H}^{(k)}\}$ are well-known in the literature, see for example \cite{Dolan:2004iy, Chicherin:2015edu}:
\begin{equation}\label{position_space_results}
    \begin{split}
       & \mathcal{H}^{(0)}(u,v)=\left(1+\frac{1}{v^2}+\frac{4}{N^2}\frac{1}{v}\right)\\ &\mathcal{H}^{(1)}(u,v)=-2\,\frac{u}{v}\,\Phi^{(1)}(u,v)\\ 
          & \mathcal{H}^{(2)}(u,v)=\frac{u}{v}\left[\frac{1}{4}(1+u+v)\Phi^{(1)}(u,v)^2+\Phi^{(2)}(u,v)+\frac{1}{v}\Phi^{(2)}\left(\frac{u}{v},\frac{1}{v}\right)+\frac{1}{u}\Phi^{(2)}\left(\frac{1}{u},\frac{v}{u}\right)\right] 
        \end{split}
    \end{equation}

Note that these are the perturbative results in  $\lambda$ in the planar (genus zero) limit. 
\section{Perturbative Mellin Amplitudes}\label{Finding Mellin Amplitudes}
In this section, we will compute the Mellin amplitudes at one and two loop orders using Mellin-Barnes representations (see Appendix \ref{Triangle Integral}) of one and two loop triangle diagrams. We will observe a stringy feature of having infinite number of poles at the two-loop Mellin amplitude, after stripping off the factor $\Gamma^2(2-s)\Gamma^2(-t)\Gamma^2(s+t)$. The latter factor accounts for all multi-trace OPE contributions corresponding to multi-string states in the dual bulk, and so the stripped off amplitude (often called `reduced Mellin amplitude') only has poles associated to single trace operators (corresponding to single-string states). 
\subsection*{One-loop Mellin amplitude}
Conformal loop integral $\Phi^{(1)}(x,y)$ is given \cite{Usyukina:1992jd} by one-loop triangle diagram  with  massless internal propagators, upto multiplicative factor (for the definition of $J(d,\nu_1,\nu_2,\nu_3)$, see appendix \ref{Triangle Integral}):
\begin{equation}
    \Phi^{(1)}(x,y)=\frac{p_3^2}{i\pi^2}J(4;1,1,1)
\end{equation}
where, $$x=\frac{p_1^2}{p_3^2}\quad\text{and}\quad y=\frac{p_2^2}{p_3^2}$$
Following \cite{Usyukina:1992jd}, Mellin-Barnes integral representation for this diagram has been found in Appendix \ref{Triangle Integral}:
\begin{equation}
 J(4;1,1,1)=-\frac{i\pi^2}{p_3^2}\int_{-i\infty}^{+\infty} \frac{dsdt}{(2\pi i)^2} \, x^s y^t\, \Gamma^2(-s)\Gamma^2(-t)\Gamma^2(1+s+t)
\end{equation}
Replacing $x=u$, $y=v$,
\begin{equation}
      \Phi^{(1)}(u,v)= \int_{-i\infty}^{+\infty} \frac{dsdt}{(2\pi i)^2}\, u^s v^t \, \Gamma^2(-s)\Gamma^2(-t)\Gamma^2(1+s+t)
\end{equation}
Hence, 
\begin{equation}
    \begin{split}
        \mathcal{H}^{(1)}(u,v)&=-2\,\frac{u}{v}\, \Phi^{(1)}(u,v)\\&=2\int_{-i\infty}^{+i\infty} \frac{dsdt}{(2\pi i)^2} \, u^s v^t\, \Gamma^2(1-s)\Gamma^2(-t-1)\Gamma^2(1+s+t)\\&=\int_{-i\infty}^{i\infty}\frac{ds}{2\pi i}\frac{dt}{2\pi i}u^sv^t\,{\Gamma^2(2-s)\Gamma^2(-t)\Gamma^2(s+t)}\, \mathcal{M}^{(1)}(s,t)
    \end{split}
\end{equation}
where,
\begin{equation}
      \mathcal{M}^{(1)}(s,t)= 2\left[\frac{1}{(s-1)^2}+\frac{2}{(s-1)(t+1)}+\frac{1}{(t+1)^2}\right]
\end{equation}
This gives the Mellin amplitude $\mathcal{M}^{(1)}(s,t)$

\subsection*{Two-loop Mellin amplitude}
For two-loop Mellin amplitude, as clear from \eqref{position_space_results}, we need to determine the Mellin-Barnes representations for  $\Phi^{(2)}(u,v)$ and $\Phi^{(1)}(u,v)^2$.
\subsubsection*{$\Phi^{(2)}(u,v)$}
The two-loop integral $\Phi^{(2)}(x,y)$ is related to two-loop triangle diagram with all internal propagators are massless:
\begin{equation}
    C^{(2)}(p_1^2,p_2^2,p_3^2)=\left(\frac{i\pi^2}{p_3^2}\right)^2\,\Phi^{(2)}(x,y)
\end{equation}
where x and y are same as defined above. \\*
Following \cite{Allendes:2012mr}, Mellin-Barnes integral representation for this diagram has been calculated in Appendix \ref{Triangle Integral} from which we get, after replacement $x=u$, $y=v$:
\begin{equation}
    \begin{split}
      \Phi^{(2)}(u,v)=\int  \frac{ds}{2\pi i} \frac{dt}{2\pi i} \, u^s v^t M^{(2)}_{\Phi^{(2)}(u,v)}(s,t)
    \end{split}
\end{equation}
where, 
\begin{equation}
    M^{(2)}_{\Phi^{(2)}(u,v)}(s,t)= \frac{1}{2}\,\Gamma^2(-s) \Gamma^2(-t) \Gamma^2(1+s+t)\,[(\Psi'(-t)+\Psi'(-s))-(\Psi(-t)-\Psi(-s))^2-\pi^2]
\end{equation}
Here $\Psi(z)$ is digamma function of $z$. After stripping off the factor $\Gamma^2(2-s)\Gamma^2(-t)\Gamma^2(s+t)$, we get the reduced Mellin amplitude as
\begin{equation}
    \mathcal{M}^{(2)}_{\Phi^{(2)}(u,v)}(s,t)=\frac{1}{2}\frac{(s+t)^2}{s^2(s-1)^2} \,[(\Psi'(-t)+\Psi'(-s))-(\Psi(-t)-\Psi(-s))^2-\pi^2]
\end{equation}
Clearly, $\mathcal{M}^{(2)}_{\Phi^{(2)}(u,v)}(s,t)$ has an infinite number of simple and double poles in $s,t$. This indicates the massive stringy excitations in the bulk. Existence of the double poles for the reduced Mellin amplitude might look suspicious at first. But we should remember, even though the non-perturbative (in $\lambda$) reduced Mellin-amplitude is a meromorphic function with only simple poles, when we make perturbation in $\lambda$, we will naturally expect to get double poles at the two-loop order. These double pole contributions can be viewed as shifts in the locations of the poles, and therefore as a measure of the anomalous dimensions. 
\subsubsection*{$[\Phi^{(1)}(u,v)]^2$}
We have the following Mellin-Barnes integral representation of $\Phi^{(1)}(u,v)$:
\begin{equation}
    \Phi^{(1)}(u,v)=-\int_{c-i\infty}^{c+i\infty}\frac{ds}{2\pi i}\frac{dt}{2\pi i}\, u^s v^t \,\Gamma^2(-s)\Gamma^2(-t)\Gamma^2(1+s+t)
\end{equation}
where real part c can be chosen to lie in the interval $(-\frac{1}{2})<c<0$. 
Taking the regime $u<1$ and $v<1$, we have to close both the contours of $s$ and $t$ on the right (i.e around the positive axis) and we will not consider any pole on the negative axes,
\begin{equation}
    \begin{split}
        \Phi^{(1)}(u,v)&=-\sum_{n=0}^{\infty} \frac{2}{(n!)^2}\,\Psi(n+1)\, u^n \int_{c-i\infty}^{c+i\infty}\frac{dt}{2\pi i}\, v^t \, \Gamma^2(-t) \Gamma^2(1+n+t)\\
        &=-\sum_{n=0}^{\infty}\sum_{m=0}^{\infty} \frac{4}{(n!)^2 (m!)^2} \Psi(n+1) \Psi(m+1) \Gamma^2 (1+n+m)\, u^n \, v^m
    \end{split}
\end{equation}
For writing a Mellin representation of $[\Phi^{(1)}(u,v)]^2$, we can use the above taylor-expansion for one $\Phi^{(1)}(u,v)$ and Mellin transform the other $[\Phi^{(1)}(u,v)]^2$, and then while multiplying we will just shift the arguments in $(s,t)$ to obtain
\begin{equation}
\begin{split}
[\Phi^{(1)}(u,v)]^2
  =\int \frac{ds}{2\pi i}\frac{dt}{2\pi i} \, u^s v^t \,  M_{\Phi^{(1)}(u,v)^2}^{(2)},
\end{split}
\end{equation}
with
\begin{equation}
    M_{\Phi^{(1)}(u,v)^2}^{(2)}=-\sum_{n=0}^{\infty}\sum_{m=0}^{\infty}\frac{4}{(n!)^2 (m!)^2} \Psi(n+1) \Psi(m+1) \Gamma^2 (1+n+m) \Gamma^2(-s+n) \Gamma^2(-t+m) \Gamma^2(1+s+t-n-m).
\end{equation}

\subsubsection*{Net two-loop Mellin amplitude:}

Collecting the above Mellin integral representations of $[\Phi^{(2)}(u,v)$ and $[\Phi^{(1)}(u,v)]^2$, we can compute the net two-loop Mellin amplitude of $\mathcal{H}^{(2)}(u,v)$ from \eqref{position_space_results}:

\begin{equation}\label{two-loopmellinamplitude}
\begin{split}
    M^{(2)}(s,t)&=\frac{1}{4}[  M_{\Phi^{(1)}(u,v)^2}^{(2)}(s-1,t+1)+  M_{\Phi^{(1)}(u,v)^2}^{(2)}(s-2,t+1)+  M_{\Phi^{(1)}(u,v)^2}^{(2)}(s-1,t)]\\&+   M^{(2)}_{\Phi^{(2)}(u,v)}(s-1,t+1)-M^{(2)}_{\Phi^{(2)}(u,v)}(s-1,-t-s-2)- M^{(2)}_{\Phi^{(2)}(u,v)}(-s-t,t+1)
    \end{split}
\end{equation}

\section{Mellin Conformal Block Expansion}\label{Mellin Conformal Block Expansion}
In this section, we will make an Mellin conformal block expansion of the amplitude, which will then be used in the next section to read off the one-loop anomalous dimensions and OPE coefficients of the exchanged twist-two singlet multiplets.

We can expand dynamic part $\mathcal{H}(u,v)$  of the correlator $\mathcal{G}(u,v;\sigma,\tau)$ in \eqref{Superconformal Ward Identity} into conformal partial waves,
\begin{equation}
   \mathcal{H}(u,v)=\sum_{t'}\sum_{l}A_{00t' l}u^{t'}G^{(l)}_{t'+2}(u,v)
\end{equation}
Here $t'=\frac{\Delta-l}{2}$ is the twist and $l$ is the spin of the exchanged operator in the OPE channel. Since operator dimension $\Delta$ receives quantum corrections, we have an expansion of $t'$ in 't Hooft coupling $\lambda$:
\begin{equation}
    t'=\frac{\Delta^0-l}{2}+\frac{1}{2}(\eta_l^{(1)}+\eta_l^{(2)}+\cdots)
\end{equation}
where $t_0'=(\Delta^0-l)/2$ is the twist in the free theory, and $\eta_l^{(k)}\sim \mathcal{O}(\lambda^k)$ is the anomalous dimension at the $k$-order in $\lambda$. 

\vspace{2mm}

Now we can taylor expand $G^{(l)}_{t'+2}(u,v)$ in $u$:
\begin{equation}
    G_{t'}^{(l)}(u,v) = \sum_{m=0}^{\infty} u^m g^{(m)}_{t',l}(v)
\end{equation}
We can write $g^{(m)}_{t',l}(v)$ in a Mellin-Barnes representation \cite{Costa:2012cb}:
\begin{equation}
    \begin{split}
        g^{(m)}_{t',l}=\frac{1}{m!}\beta_{l,m}(t')\int \frac{dt}{2\pi i}\,v^t\,Q_{l,m}^{t'}(t)\, \Gamma^2(-t)\Gamma^2(t+t'+m)
    \end{split}
\end{equation}
where, 
\begin{equation}
    \beta_{l,m}(t')=\frac{\Gamma(2t'+2l)(2t'+l-1)_l}{4^l(2t'+l-h+1)_m\Gamma^{4}(t'+l)}
\end{equation}
Note, our definition of $Q_{l,m}^{t'}(t)$ differs \footnote{In fact, $Q_{J,m}^{t'}(t)=Q_{J,m}(-2(t+t'+m))$ where $Q_{J,m}$ is what is defined in \cite{Costa:2012cb}.} from that in \cite{Costa:2012cb}. $Q^{t'}_{l,m}(t)$ are related to the Mack polynomials $P^{(s)}_{\nu,l}(s,t)$ (see \cite{Gopakumar:2018xqi} for it's defintion) as\footnote{$h=\frac{d}{2}$, $d$ being space-time dimension of the CFT, i.e $d=4$ here.}:
\begin{equation}
    Q^{t'}_{l,m}(t) = \dfrac{4^l}{(2t'+l-1)_l(2h-t'-l-1)_l}P^{(s)}_{2t'+l-h,l}(s = t'+ m,t)
\end{equation}
In particular, $Q^{t'}_{l,0}(t)$ turns out to be continuous Hahn polynomials $Q^{t'}_{l,0}(t)$ written as
\begin{equation}
    Q^{t'}_{l,0}(t) = \dfrac{2^l(t')^2_l}{(2t'+l-1)_l} {}_3 F_2[-l,2t'+l-1,t'+t;t',t';1]
\end{equation}
and they satisfy an orthogonality relation
\begin{equation}\label{orthogonality}
    \int_{-i\infty}^{i\infty}\dfrac{dt}{2\pi i} \Gamma^2(t + t')\Gamma^2(-t) \, Q^{t'}_{l,0}(t)Q^{t'}_{l',0}(t) = \kappa_l(t')\,\delta_{l,l'}
\end{equation}
with the coefficients $\kappa_l$ given by:
\begin{equation}
     \kappa_l(t') = \dfrac{(-4)^l l!}{(2t'+l-1)^2_l}\dfrac{\Gamma^4(t'+l)}{(2t'+ 2l-1)\Gamma(2t'+l-1)}
\end{equation}
This orthogonality relation would play a crucial role in reading off the anomalous dimension and OPE coeffcients in the next section. Combining these results, we can find a conformal block-like expansion in mellin space of the Mellin amplitudes. We rewrite position space expansion as Mellin integral:
\begin{equation}
    \begin{split}
       \mathcal{H}(u,v)&=\sum_{\Delta^{0},l}A^{t'}_{l}u^{t'}\left[\sum_{m=0}^{\infty}u^m g^{(m)}_{t'+2}(v)\right]\\&=\sum_{\Delta^{0},l}A^{t'}_{l}u^{t'}\sum_{m=0}^{\infty}\frac{1}{m!}u^m \beta_{l,m}(t'+2)\int_{-i\infty}^{i\infty}\frac{dt}{2\pi i}v^t \Gamma^2(t+t'+2+m)\Gamma^2(-t)Q^{t'+2}_{l,m}(t)\\&=\sum_{\Delta^{0},l}A^{t'}_{l}\sum_{m=0}^{\infty} \frac{1}{m!}\beta_{l,m}(t'+2)\int_{-i\infty}^{i\infty}\frac{ds}{2\pi i}\frac{dt}{2\pi i}u^sv^t\Gamma^2(t+t'+2+m)\Gamma^2(-t)\frac{Q^{t'+2}_{l,m}(t)}{s-(t'+m)}
    \end{split}
\end{equation}
Since $\mathcal{H}(u,v)$ has the standard Mellin integral representation
\begin{equation}
    \mathcal{H}(u,v)=\int_{-i\infty}^{i\infty}\frac{ds}{2\pi i}\frac{dt}{2\pi i}\, u^sv^t\,M(s,t),
\end{equation}
we can write an expansion of the Mellin amplitude $M(s,t)$: 
\begin{equation}\label{Conformal Block Expansion}
\boxed{
M(s,t)=\sum_{\Delta^{0},l}A^{t'}_{l} \beta_{l,m}(t'+2)\Gamma^2(t+t'+2+m)\Gamma^2(-t)\sum_{m=0}^{\infty}\frac{1}{m!}\frac{Q^{t'+2}_{l,m}(t)}{s-(t'+m)}}
\end{equation}

\section{Anomalous Dimension and OPE Coefficient at one-loop}
To read off the anomalous dimension $\eta_l$ and OPE coefficient from \eqref{Conformal Block Expansion}, we take a particular ``kinematic" limit: $s \rightarrow 1$ on both sides of the relation. Then clearly the term that dominates on the sum in r.h.s comes from with $t'_0+m=1$, and this has has unique solution $t'_0=1$ and $m=0$ coming from unitary constraints of the CFT. 

We can thus express
\begin{equation}\label{s to 1 limit}
    \begin{split}
       M(s\rightarrow 1,t)=\sum_{l}\frac{A_{l}^{t'} \, Y_{l}(t',t)}{s-t'};\quad t'=1+\frac{\eta_l}{2}
    \end{split}
\end{equation}
where $\beta_{l,0}(t')\equiv \beta_{l}(t')$ and $ Y_{l}(t',t)$ is defined as:
\begin{equation}
    \begin{split}
         Y_{l}(t',t)=\beta_{l}(t'+2)\Gamma^2(t+t'+2)\Gamma^2(-t)Q^{t'+2}_{l,0}(t)
    \end{split}
\end{equation}
We can perform a 't Hooft coupling expansion of each sub-pieces on the r.h.s of \eqref{s to 1 limit}:
\begin{equation}
    \begin{split}
        A^{t'}_{l}&=A_{l,0}(1+\lambda\, b_{l,1}+\lambda b_{l,2})\\
        Y_{l}(t',t)&=Y_{l}(1,t)+\lambda \frac{1}{2}\eta_{l,1}Y'_{l}(1,t)+\lambda^2\frac{1}{2}\eta_{l,2} Y'_{l}(1,t)+\lambda^2 \frac{1}{8} \eta_{l,1}^2 Y''_{l}(1,t)\\ 
        \frac{1}{s-t'}&=\frac{1}{s-1}+\lambda \frac{\eta_{l,1}/2}{(s-1)^2}+\lambda^2 \frac{\eta_{l,2}/2}{(s-1)^2}+\lambda^2\frac{\eta_{l,1}^2/4}{(s-1)^3}
    \end{split}
\end{equation}
Putting these together in \ref{s to 1 limit}, we get
\begin{equation}\label{perturbativeconformalblockexpansion}
    \begin{split}
        M^{(0)}(s\to 1,t)&=\sum_{l} A_{l,0}\frac{Y_{l}(1,t)}{s-1}\\
         M^{(1)}(s\to 1,t)&=\sum_{l} A_{l,0}\Big[ \frac{b_{l,1}Y_l(1,t)+\frac{1}{2}\eta_{l,1}Y'_{l}(1,t)}{s-1}+\frac{\frac{1}{2}\eta_{l,1}Y_{l}(1,t)}{(s-1)^2}\Big]\\
         M^{(2)}(s\to 1,t)&=\sum_{l} A_{l,0}\Big[ \frac{b_{l,2}\,Y_{l}(1,t)+\frac{1}{2}\eta_{l,2}Y'_{l}(1,t)+\frac{1}{8}\eta_{l,1}^2 Y''_{l}(1,t)+\frac{1}{2}\eta_{l,1}b_{l,1}Y'_{l}(1,t)}{s-1}\\&+\frac{\eta_{l,2}Y_{l}(1,t)+\frac{1}{4}\eta_{l,1}^2 Y'_{l}(1,t)+\frac{1}{2}\eta_{l,1}b_{l,1}Y_{l}(1,t)}{(s-1)^2}+\frac{\frac{1}{4}\eta_{l,1}^2 Y_{l}(1,t)}{(s-1)^3}\Big]
    \end{split}
\end{equation}
\subsubsection*{The strategy}\footnote{Here we will show the strategy to determine the anomalous dimension $\eta_{l,2}$ coming from $\Phi^{(2)}$ pieces of the full two-loop Mellin amplitude \eqref{two-loopmellinamplitude}: 
\begin{equation}\label{Phi-2contribution}
      M^{(2)}_2(s,t)=M^{(2)}_{\Phi^{(2)}(u,v)}(s-1,t+1)-M^{(2)}_{\Phi^{(2)}(u,v)}(s-1,-t-s-2)- M^{(2)}_{\Phi^{(2)}(u,v)}(-s-t,t+1)
\end{equation}
It is important to note that $M^{(2)}_{\Phi^{(2)}(u,v)}(s,t)$ has poles around $s=1$ upto fourth-order $\frac{1}{(s-1)^4}$. It might seem incompatible with (\ref{perturbativeconformalblockexpansion}) where we have maximum 3-rd order pole around $s=1$. But once we sum up the three terms in (\ref{Phi-2contribution}), the fourth order poles get cancelled!

Considering the 2-nd order pole in (\ref{Phi-2contribution}) and matching it with that in \eqref{perturbativeconformalblockexpansion},
\begin{equation}
    \sum_{l} A_{l,0}(\eta_{l,2}+\frac{1}{2}\eta_{l,1}b_{l,1})\beta_{l}(3)\Gamma^2(t+3) Q^{3}_{l,0}(t)= G(t) - \frac{1}{4}\sum_{l} A_{l,0} \eta_{l,1}^2 \frac{\partial}{
    \partial t'}[ \beta_{l}(t'+2)\Gamma^2(t+t'+2)Q^{t'+2}_{l,0}(t)]
\end{equation}
 with,
\begin{equation}
    G(t)=\Gamma^2(1+t) \Big[ -\frac{1}{(t+2)^2}-\Gamma^2(-t)\Gamma^2(1+t)-h(-t-2)^2+4 \cos{(\pi t)} \Gamma(-t)\Gamma(1+t) h(t+1)+h(t+2)^2\Big]
\end{equation} 
It should be now straightforward to find $\eta_{l,2}$ from the above equation.}
We have full one-loop amplitude in position space as:
\begin{equation}
    \begin{split}
     M^{(1)}(s,t)=2\,\Gamma^2(1-s)\Gamma^2(-t-1)\Gamma^{2}(1+s+t)
    \end{split}
\end{equation}
Taking $s\rightarrow 1$ limit,
\begin{equation}\label{eq:20}
    \begin{split}
      M^{(1)}(s\rightarrow 1,t)&=
     2\, \left[ \frac{1}{(s-1)^2}+ \frac{2\gamma}{s-1}\right]\Gamma^2(-t-1) \Gamma^2(2+t)\left[ 1+2(s-1)\Psi(2+t)\right]\\&= 2\, \Gamma^2(-t)\Gamma^2(1+t) \left[ \frac{1}{(s-1)^2}+\frac{2 h(1+t)}{s-1}\right]
    \end{split}
\end{equation}
Matching the residues of the simple and double poles at $s=1$ on both sides of (\ref{perturbativeconformalblockexpansion}) for one-loop, we get
\begin{equation}\label{anomalous dimension eq}
\boxed{
    \sum_{l}\eta_{l,1}\, \widetilde{A}_{l,0}\,\beta_{l}(3)\,\frac{\Gamma^2(t+3)}{\Gamma^2(1+t)}\,Q^{3}_{l,0}(t)=1}
\end{equation}
and 
\begin{equation}\label{coefficient equation}
\boxed{
    \sum_{l}\tilde{A}_{l,0}b_{l}\,\beta_{l}(3)\,\frac{\Gamma^2(t+3)}{\Gamma^2(1+t)}Q^{3}_{l,0}(t)+\sum_{l}\frac{\eta_{l,1}}{2}\tilde{A}_{l,0}\frac{\partial}{\partial t'}[\beta_{l}(t'+2)\frac{\Gamma^2(t+t'+2)}{\Gamma^2(1+t)}Q^{t'+2}_{l,0}(t)]|_{t'=1} = h(1+t)}
\end{equation}
where, $A_{l,0}=4\,\widetilde{A}_{l,0}$.

\subsection*{Anomalous dimension at one-loop}
From the the expression \eqref{anomalous dimension eq}, we can determine one-loop anomalous dimension $\eta^{1}_l$ using the orthogonality relation (\ref{orthogonality}):

\begin{equation}\label{Anomalous dimension}
    \begin{split}
        \widetilde{A}_{l,0}\eta_{l,1}&=\frac{1}{\beta_{l}(3)\kappa_{l}(3)}\Big[\int_{c-i\infty}^{c+i\infty}\dfrac{dt}{2\pi i}   \Gamma^2(-t)\Gamma^2(1+t) Q^{3}_{l,0}(t)\Big] \\
    &= \dfrac{2^l\Gamma^2(l+3)\Gamma(l+5)}{(2!)^2 l! \Gamma(2l+5)} \Big[\int_{c-i\infty}^{c+i\infty}\dfrac{dt}{2\pi i}   \Gamma^2(-t)\Gamma^2(1+t) {}_3 F_2[-l,l+5,3+t;3,3;1]\Big]
    \end{split}
\end{equation}
where $-2 < c < -1$.

Using the integral representation,
\begin{equation}\label{3F2}
{}_3 F_2[-l,l+5,3+t;3,3;1] = \frac{\Gamma(3)}{\Gamma(3+t)\Gamma(-t)} \int_{0}^{1}dw w^{t+2}(1-w)^{-t-1}{}_2 F_1[-l,l+5;3;w]
\end{equation}
we can rewrite as,
\begin{equation}
    \begin{split}
       \widetilde{A}_{l,0}\eta_{l,1}&=  \big[\dfrac{2^l\Gamma^2(l+3)\Gamma(l+5)\Gamma(3)}{(2!)^2 l! \Gamma(2l+5)}\Big] \int_{0}^{1} dw {}_2 F_1[-l,l+5;3;w] \int_{c-i\infty}^{c+i\infty} [dt] w^{t+2} (1-w)^{-t-1} \frac{\Gamma(-t)\Gamma^{2}(1+t)}{\Gamma(3+t)}
    \end{split}
\end{equation}
We can find the last integral explicitly as,
\begin{equation}
    \int_{c-i\infty}^{c+i\infty} [dt] w^{t+2} (1-w)^{-t-1} \frac{\Gamma(-t)\Gamma^{2}(1+t)}{\Gamma(3+t)}=-[w\ln{w}+(1-w)\ln{(1-w)}]
\end{equation}
Using this and the followings,
\begin{equation}
    {}_2 F_1[-l,l+5;3;w]=\sum_{k=0}^{\infty} \frac{(-l)_k (l+5)_k}{(3)_k\, k!} w^k;\quad |w|<1.
\end{equation}
and 
\begin{equation}
    [-\int_{0}^{1} dw \, w^k  [w\ln{w}+(1-w)\ln{(1-w)}]=\frac{h(1+k)}{(k+1)(k+2)}
\end{equation}
We finally have
\begin{equation}\label{dolan-1}
\boxed{
  \eta_l^{(1)} = \dfrac{\Gamma(l+5)}{2l!}\sum_{k=0}^{\infty}\dfrac{(-l)_k (l+5)_k}{\Gamma^2(k+3)}h(k+1) = \begin{cases}
       2h(l+2) &\quad l = 0,2,4,.. \\
       0 &\quad l = 1,3,5,.. \end{cases}}
\end{equation}
This coincides with the result Eq. (5.24) of Dolan, Osborn \cite{Dolan:2004iy}.

\subsection*{OPE coefficient at one-loop}
We broke the expression \ref{coefficient equation} into two steps:
\begin{equation}\label{1st}
\sum_{l}\widetilde{A}_{l,0}b^{(1)}_{l}[\beta_{l}(3)\frac{\Gamma^2(t+3)}{\Gamma^2(1+t)}Q^{3}_{l,0}(t)]=   h(1+t)
\end{equation}

\begin{equation}\label{2nd}
\sum_{l}\widetilde{A}_{l,0}b^{(2)}_{l}[\beta_{l}(3)\frac{\Gamma^2(t+3)}{\Gamma^2(1+t)}Q^{3}_{l,0}(t)]+\sum_{l}\frac{\eta^{(1)}_l}{2}\widetilde{A}_{l,0}\frac{\partial}{\partial t'}[\beta_{l}(t'+2)\frac{\Gamma^2(t+t'+2)}{\Gamma^2(1+t)}Q^{t'+2}_{l,0}(t)]|_{t'=1}=0
\end{equation}
with $b^{(1)}_{l}+b^{(2)}_{l}=b_{l}$.\\*
Eq. (\ref{1st}) can be solved for $b^{(1)}_l$:
\begin{equation}\label{bl}
\begin{split}
    b^{(1)}_l &= \dfrac{(-4)^l \Gamma^4(3 +l)}{\Gamma(6 + 2l)(6 + l -1)_l} [\kappa_l(3)]^{-1} \Big[\int_{-i\infty}^{i\infty}\dfrac{dt}{2\pi i}   \Gamma^2(-t)\Gamma^2(1+t)h(1+t) Q^{3}_{l,0}(t)\Big] \\
    &= \dfrac{2^l\Gamma^2(l+3)\Gamma(l+5)}{(2!)^2 l! \Gamma(2l+5)} \Big[\int_{-i\infty}^{i\infty}\dfrac{dt}{2\pi i}   \Gamma^2(-t)\Gamma^2(1+t)h(1+t) {}_3 F_2[-l,l+5,3+t;3,3;1]\Big]
    \end{split}
\end{equation}   
Using the integral representation (\ref{3F2}) for ${}_3 F_2$ and the following integral representation for $h(z)$,
\begin{equation}
    h(z) = \int_0^1 \dfrac{1-t^z}{1-t}dt
\end{equation}
Eq. (\ref{bl}) can be simplified to obtain 
\begin{align}
     b^{(1)}_l &= \dfrac{\Gamma(l+5)}{2^2 l!}\sum_{s=0}^{l}\dfrac{(-l)_s (l+5)_s}{\Gamma^2(s+3)}(-h^{(2)}(s+1) - h^2(s+1))\\
           &= (-1)^{l+1} h(l+2)^2 + \sum_{0}^{l+2}(-1)^r\dfrac{1}{r^2}
\end{align}

To compute  $b^{(2)}_l$, we have to consider the derivative $\dfrac{\partial}{\partial \Delta} g^{(0)}_{t',l}(1-v)$. This gives us the equation
\begin{equation}
\sum_{l}A_{1,l} \Delta^{(1)}_l \dfrac{\partial}{\partial \Delta}g_{t^{'}+2,l}(1-v)|_{t^{'} =1} = - \sum_{l}A_{1,l} b_l^{(2)}g_{3,l}(1-v)
\end{equation}
In Mellin space, the equation becomes
\begin{equation}
\begin{split}
 \sum_l A_{1,l}  \dfrac{\Delta^{(1)}_l}{2}  \dfrac{\partial}{\partial t^{'}}\Big[\beta_l(t') \Gamma^2(-t)\Gamma^2(t+t^{'}+2)Q^{t^{'}+2}_l(t)\Big]_{t^{'}=1} \\ = -\sum_lb_l^{(2)} A_{1,l}  \Big[\beta_l(t') \Gamma^2(-t)\Gamma^2(t+t^{'}+2)Q^{t^{'}+2}_l(t)\Big]_{t^{'}=1}    
\end{split}
\end{equation}
Using orthogonality, we get
\begin{equation}
\sum_l A_{1,l}  \dfrac{\Delta^{(1)}_l}{2}  \dfrac{\partial}{\partial t^{'}}\Big[\beta_l(t'+2) \Gamma^2(-t)\Gamma^2(t+t^{'}+2)Q^{t^{'}+2}_l(t)\Big]_{t^{'}=1}Q^{t^{'}+2}_{l^{'}}(t)|_{t^{'}=1} = -b_{l{'}}^{(2)} A_{1,l'}  l'!
\end{equation}
Now we'll use the chain rule to rewrite this equation as
\begin{multline}
\sum_l A_{1,l}  \dfrac{\Delta^{(1)}_l}{2}  \dfrac{\partial}{\partial t^{'}}\Big[\beta_l(t'+2) \Gamma^2(-t)\Gamma^2(t+t^{'}+2)Q^{t^{'}+2}_l(t)Q^{t^{'}+2}_{l^{'}}(t)\Big]_{t^{'}=1} \\
- \sum_l A_{1,l}  \dfrac{\Delta^{(1)}_l}{2}  \Big[\beta_l(t'+2) \Gamma^2(-t)\Gamma^2(t+t^{'}+2)Q^{t^{'}+2}_l(t)\Big]_{t^{'}=1}\dfrac{\partial}{\partial t^{'}}Q^{t^{'}+2}_{l^{'}}(t)|_{t^{'}=1}
= -b_{l{'}}^{(2)}A_{1,l'}  l{'}!    
\end{multline}
Now using orthogonality in the first term on rhs, the first term vanishes and for the second term, we use the result (5.22) to get
\begin{equation}
b_{l}^{(2)} = \dfrac{1}{A_{1,l} l!}\int_{\gamma -i\infty}^{\gamma + i\infty}\dfrac{dt}{2\pi i}\dfrac{1}{2} \Gamma^2(-t)\Gamma^2(1+t)\dfrac{\partial}{\partial t^{'}}Q^{t^{'+2}}_l(t)|_{t^{'}=1}
\end{equation}   
where $-2 < \gamma < -1$.\\
Computing this integral involves a lengthy calculation. It goes as follows.\\
\begin{align}
&\dfrac{\partial}{\partial t^{'}}Q^{t^{'}+2}_l(t)|_{t^{'}=1} = \dfrac{\partial}{\partial t^{'}} \Big[\frac{2^l (t' +2)_l^2}{(2t'+l+3)_l} {}_3F_2[-l,2t'+l+3, t+t'+2;t'+2,t'+2;1] \Big]_{t'=1} \\
&=  \dfrac{\partial}{\partial t^{'}} \Big[\frac{2^l (t' +2)_l^2}{(2t'+l+3)_l} \frac{\Gamma(t'+2)}{\Gamma(t+t'+2)\Gamma(-t)} \int_0^1 dr r^{t+t'+1} (1-r)^{-t-1} {}_2F_{1} [ -l, 2t'+l+3; t'+2;r] \Big]_{t'=1}
\end{align}
where we used the integral representation of ${}_3F_{2}$ in terms of ${}_2F_1$. Then we use the series expansion of ${}_2F_1$ to write \\
\begin{align}
&= \dfrac{\partial}{\partial t^{'}} \Big[\frac{2^l (t' +2)_l^2}{(2t'+l+3)_l} \frac{\Gamma(t'+2)}{\Gamma(t+t'+2)\Gamma(-t)} \int_0^1 dr r^{t'} \Big(\frac{r}{1-r}\Big)^{t+1} \sum_l \frac{(-l)_s (2t' +l +3)_s}{(t'+2)_s s! }r^s \Big]_{t'=1} \\
&=\Big [\frac{2^l (t' +2)_l^2}{(2t'+l+3)_l} \frac{\Gamma(t'+2)}{\Gamma(t+t'+2)\Gamma(-t)} \int_0^1 dr r^{t'} \Big(\frac{r}{1-r}\Big)^{t+1} \sum_l \frac{(-l)_s (2t' +l +3)_s}{(t'+2)_s s! }r^s \\
& \Big(2 h(t'+l+1) - 2h(2t' + 2l +2) - h(t' + s +1) + 2h(2t'+l+s +2) - h(t+t'+1) + \text{log }r \Big) \Big]_{t'=1} \\
& = \frac{2^l (3)_l^2}{(l+5)_l} \frac{\Gamma(3)}{\Gamma(t+3)\Gamma(-t)} \int_0^1 dr r \Big(\frac{r}{1-r}\Big)^{t+1} \sum_l \frac{(-l)_s (l +5)_s}{(3)_s s! }r^s \\
& \Big(2 h(l+2) - 2h(2l +4) - h(s +2) + 2h(l+s +4) - h(t+2) + \text{log }r \Big)
\end{align}
This gives us
\begin{align}
b_{l}^{(2)} A_{1,l} l! = \frac{1}{2}\Big[\frac{2^l (3)_l^2\Gamma(3)}{(l+5)_l} \Big]\sum_l \frac{(-l)_s (l +5)_s}{(3)_s s! } \int_0^1 dr r^s \Big[r \int \frac{dt}{2 \pi i} \frac{\Gamma(-t)\Gamma(1+t)}{(t+1) (t+2)} \Big(\frac{r}{1-r}\Big)^{t+1} \\
 \Big(2 h(l+2) - 2h(2l +4) - h(s +2) + 2h(l+s +4) - h(t+2) + \text{log }r \Big)\Big]
\end{align}
Lets first simply the part in the long brackets involving the $t$ integral. We can shift the contour to write 
\begin{align}
r \int_{\gamma - i \infty}^{\gamma+ i \infty} \frac{dt}{2 \pi i} \frac{\Gamma(1-t)\Gamma(t)}{(t) (t+1)} \Big(\frac{r}{1-r}\Big)^{t} \Big(2 h(l+2) - 2h(2l +4) - h(s +2) + 2h(l+s +4) - h(t+1) + \text{log }r \Big)
\end{align}
where $-1 < \gamma < 0$.
To perform the $t$ integral, we need the following result:
\begin{equation}
r \int_{\gamma - i \infty}^{\gamma+ i \infty} \frac{dt}{2 \pi i} \frac{\Gamma(1-t)\Gamma(t)}{(t) (t+1)} \Big(\frac{r}{1-r}\Big)^{t}  = -[ (1-r) \text{log} (1-r) + r \text{log} (r)]
\end{equation}
and its generalization
\begin{equation}
r p\int_{\gamma - i \infty}^{\gamma+ i \infty} \frac{dt}{2 \pi i} \frac{\Gamma(1-t)\Gamma(t)}{(t) (t+1)} \Big(\frac{r p}{1-r}\Big)^{t}  = r p \text{log} \Big(\frac{1-r}{r p}\Big) + (1-r+rp) \text{log} \Big(\frac{1-r+rp}{1-r}\Big)
\end{equation}
Using this, the part without $h(t+1)$ equals
\begin{equation}
-[ (1-r) \text{log} (1-r) + r \text{log} (r)]\Big(2 h(l+2) - 2h(2l +4) - h(s +2) + 2h(l+s +4) + \text{log }r \Big)
\end{equation}
And to compute the part with $h(t+1)$, we use its integral representation to get
\begin{align}
r \int \frac{dt}{2 \pi i} \frac{\Gamma(1-t)\Gamma(t)}{(t) (t+1)} \Big(\frac{r}{1-r}\Big)^{t} \int_0^1 dp \frac{(1-p^{t+1})}{1- p}
\end{align}
and then using the above results to do the t-integral, we get
\begin{equation}
\int_0^1 dp \Big[ -r \text{log}r + \frac{r p \text{log}p - (1-r +rp) \text{log}(1-r+ rp)}{1-p}\Big]
\end{equation}
The $p$ integral can be solved to get
\begin{equation}
 -\Big[ r \text{log}r + (1-r) \text{log} (1-r) \Big] - \frac{\pi^2}{6}r + \text{Li}_2(r)
\end{equation}
Now we only have the $r$ integral left to do. For that we use the results
\begin{align}
\int_0^1 dr r^s \Big[(1-r) \text{log} (1-r) + r \text{log}r\Big] &= \frac{-h(s+1)}{(s+1)(s+2)} \\
\int_0^1 dr r^s \Big[(1-r) \text{log} (1-r) + r \text{log}r\Big]\text{log} r &= \frac{1}{(s+1)^2(s+2)} + \frac{(2s+3)h(s+1)}{(s+1)^2 (s+2)^2} - \frac{\psi'(s+1)}{(s+1)(s+2)} \\
\int_0^1 dr r^s \text{Li}_2(r) &= \frac{\pi^2}{6}\frac{1}{(s+1)} - \frac{h(s+1)}{(s+1)^2}
\end{align}
Putting these all together, (13) can be written as
\begin{multline}
b_{l}^{(2)} = \frac{1}{2 A_{1,l} l! }\frac{2^l (3)_l^2\Gamma(3)}{(l+5)_l} \sum_l \frac{(-l)_s (l +5)_s}{(3)_s s! } \Big[ \frac{h(s+1)}{(s+1)(s+2)} \Big(2 h(l+2) - 2h(2l+4) \Big) \\
+ \frac{h(s+1)}{(s+1)(s+2)}2h(l+s+4) - \frac{h(s+1)h(s+2)}{(s+1)(s+2)} + \Big( \frac{\psi'(s+1)}{(s+1)(s+2)} - \frac{\pi^2}{6(s+1)(s+2)}\Big) \\
- \frac{h(s+1)}{(s+1)(s+2)} - \frac{1}{(s+1)^2(s+2)} +\Big(\frac{h(s+1)}{(s+1)^2} - \frac{(2s+3)h(s+1)}{(s+1)^2(s+2)^2}\Big)\Big]
\end{multline}
Now the prefactor outside the long brackets can be simplified using the expression for $A_{1,l}$ to get
\begin{equation}
\frac{1}{2 A_{1,l} l! }\frac{2^l (3)_l^2\Gamma(3)}{(l+5)_l} \sum_l \frac{(-l)_s (l +5)_s}{(3)_s s! } = \frac{1}{4 l!}\sum_l \frac{(-l)_s \Gamma(l+s+5)}{(s+2)! s!}
\end{equation}
and the part inside long brackets can be further simplified to get
\begin{equation}
\frac{h(s+1)}{(s+1)(s+2)}[ 2h(l+2) - 2h(2l+4)] - \frac{h(s+1)^2}{(s+1)(s+2)} - \frac{h^{(2)}(s+1)}{(s+1)(s+2)} + \frac{2 h(s+1)h(l+s+4)}{(s+1)(s+2)} - \frac{2h(s+1)}{(s+1)(s+2)^2}
\end{equation}
where we've used the relation $\psi'(s+1) =  \frac{\pi^2}{6} -  h^{(2)}(s)$ to get rid of the polygamma function.\\
Putting (26) and (27) together gives
\begin{align}
b_{l}^{(2)} &= \dfrac{\Gamma(l+5)}{4l!}\sum_{s=0}^{l}\dfrac{(-l)_s (l+5)_s}{\Gamma^2(s+3)}\Big[h(s+1)[2h(l+2) - 2h(l+4)] \\ 
&+ [-h^{(2)}(s+1)^2 - h^2(s+1)] + 2 h(s+1)[h(l+s+4) - \dfrac{1}{s+2}]\Big]    
\end{align}
which using the following identity
\begin{equation}
\begin{split}
\dfrac{\Gamma(l+5)}{2l!}\sum_{s=0}^{l}\dfrac{(-l)_s (l+5)_s}{\Gamma^2(s+3)}h(s+1)(h(l+s+4) - \dfrac{1}{s+2}) \\ =  \begin{cases}
       2h(l+2)^2 - h^{(2)}(l+2) - 2 \sum_{r=1}^{l+2} \dfrac{(-1)^r}{r^2}&\quad l = 0,2,4,.. \\
       -2h(l+2)^2 - 2 \sum_{r=1}^{l+2} \dfrac{(-1)^r}{r^2} &\quad l = 1,3,5,..
        \end{cases}     
\end{split}
\end{equation}
gives
\begin{equation}
 b_{l}^{(2)} =    \begin{cases}
       3h(l+2)^2 - 2h(l+2)h(2l+4) -h^{(2)}(l+2) - \sum_{r=1}^{l+2} \dfrac{(-1)^r}{r^2}&\quad l = 0,2,4,.. \\
       -h(l+2)^2 - \sum_{r=1}^{l+2} \dfrac{(-1)^r}{r^2} &\quad l = 1,3,5,..
\end{cases} 
\end{equation}
Therefore the coefficient correction $B^1_{00,l} = b^{(1)}_l + b^{(2)}_l$ becomes
\begin{equation}\label{dolan-2}
\boxed{
B_{1,l} =    \begin{cases}
       2h(l+2)^2 - 2h(l+2)h(2l+4) -h^{(2)}(l+2) &\quad l = 0,2,4,.. \\
       0 &\quad l = 1,3,5,..
\end{cases} }
\end{equation}
These results of anaomalous dimension and coefficent correction at one-loop in (\ref{dolan-1}) and (\ref{dolan-2}) coincide precisely with those found in \cite{Dolan:2004iy} using position-space correlators. Our analysis is rather straightforward for utilising the orthogonality relation \eqref{orthogonality} of continuous Hahn polynomials. The straightforwardness might be indicative of naturalness of Mellin space to formulated CFTs in general.

\section*{Acknowledgements}
We thank Rajesh Gopakumar, R. Loganayagam, and Aninda Sinha for useful discussions during the course of this project. We especially thank Rajesh Gopakumar for suggesting this problem to us, his guidance throughout this work, and his comments on the draft. Bulk of this material is based on Master's theses of the authors done at ICTS-TIFR. F.B is supported by the Prime Minister's Research Fellowship (PMRF). P.M is supported is supported by the Swiss National Science Foundation. 

\appendix

\section{Comments on bootstrapping one-loop amplitude}
In this section, we want to show that if we assume certain asymptotic growth of our Mellin amplitude at one-loop, then it is actually a unique solution to a algebraic problem, very similar to what has been put forward by Rastelli-Zhou \cite{Rastelli:2017udc} for tree-level supergravity amplitudes. Before that, we rewrite our one-loop mellin amplitude with the follwoing definition for Mellin integral:
\begin{equation}
    \begin{split}
         \mathcal{H}^{(1)}(u,v)&=\int_{-i\infty}^{i\infty}\frac{ds}{2\pi i}\frac{dt}{2\pi i}u^{s/2}v^{t/2-2}{\Gamma^2(1-s/2)\Gamma^2(1-t/2)\Gamma^2(1-\widetilde{u}/2)}\widetilde{\mathcal{M}}^{\text{one-loop}}(s,t)
    \end{split}
\end{equation}
where, $\widetilde{u}=4-s-t$. Then we have,
\begin{equation}\label{Rastelli-way}
    \widetilde{\mathcal{M}}^{\text{one-loop}}(s,t)=\frac{32}{(s-2)^2(t-2)^2(\widetilde{u}-2)^2}
\end{equation}
This manifests crossing symmetry, i.e, invariance under permutation of $s,t,\widetilde{u}$ (which follows from OPE associativity). Interestingly, the pole structure is exactly square of the tree-level supergravity amplitude (in Mellin space) found in \cite{Rastelli:2017udc, Alday:2018kkw}:
\begin{equation}
     \widetilde{\mathcal{M}}^{\text{sugra}}(s,t)\sim\frac{1}{(s-2)(t-2)(\widetilde{u}-2)}
\end{equation}
with
\begin{equation}
\widetilde{\mathcal{M}}^{\text{one-loop}}(s,t)\sim   [\widetilde{\mathcal{M}}^{sugra}(s,t)]^2
\end{equation}
In the following we mention the following bootstrap criteria, which could give our one-loop Mellin amplitude as a solution:
\begin{enumerate}
    \item \textbf{Superconformal Symmetry in Mellin space:}  Superconformal ward identity amounts to the following relation for Mellin amplitudes \cite{Rastelli:2017udc}:
    \begin{equation}
        \mathds{M}(s,t;\sigma,\tau)=\widehat{\mathcal{S}} \circ \widetilde{\mathcal{M}}(s,t)
    \end{equation}
    where $\widehat{\mathcal{S}}$ is the difference operator \cite{Rastelli:2017udc} corresponding to $S(u,v,\sigma,\tau)$ prefactor in Eq. (\ref{Superconformal Ward Identity}), $\mathds{M}(s,t;\sigma,\tau)$ and $\mathcal{M}(s,t)$ are reduced Mellin amplitudes of $\mathcal{G}(u,v'\sigma,\tau)$ and $\mathcal{H}(u,v)$ respectively. In particular it acts on $\widetilde{\mathcal{M}}(s,t)$ in the following way:
    \begin{equation}
        \begin{split}
           \widehat{u^mv^n}\circ \widetilde{\mathcal{M}}(s,t)\equiv \widetilde{\mathcal{M}}(s-2m,t-2n;\sigma,\tau)\, \left(2-\frac{s}{2}\right)_{m}^2\left(2-\frac{t}{2}\right)_{n}^2\left(2-\frac{u}{2}\right)_{2-m-n}^2
        \end{split}
    \end{equation}
    where, $u=8-s-t$ and $(a)_n$ as uaual Pochhammer symbol.
    \item \textbf{Crossing Symmetry:} It implies
    \begin{equation}
        \widetilde{\mathcal{M}}(s,t)=\widetilde{\mathcal{M}}(t,s)=\widetilde{\mathcal{M}}(\widetilde{u},t)
    \end{equation}
    \item \textbf{Asympotic Behaviour:} 
    \begin{equation}
       \begin{split}
          \mathds{M}(\beta s,\beta t) \sim \mathcal{O}\left(\frac{1}{\beta^2}\right)\,\Rightarrow \,
        \widetilde{\mathcal{M}}(\beta s,\beta t)\sim \mathcal{O}\left(\frac{1}{\beta^6}\right)\quad \text{for}\quad \beta \rightarrow \infty
        \end{split}
        \end{equation}
        We don't have any a priori explanation for this asympotic behaviour. For supergravity case, \cite{Rastelli:2017udc}, argued their $\mathcal{O}(\beta)$ growth using Penedones' flat-space conjecture \cite{Penedones:2010ue}. In particular, since contact interactions with $2n$ derivatives give a power n growth and IIB supergravity (in ten-dimesnional flat space) contains contact interactions with at most two derivatives, corresponding Mellin amplitude should have power-one $\mathcal{O}(\beta)$ growth. 
        \item \textbf{Analytic Structure:}
   Since $\Gamma$-stripped off mellin amplitude $M(s,t)$ has poles corresponding to twists of the exchanged single-trace operators in the OPE at 
   $$ s=\tau+2n $$
   where $\tau=2\quad \text{and}\quad n=0,1,2,...$. At one-loop ($O(\lambda)$), we only have poles at $s=2$.
\end{enumerate}
The unique solution to this algebraic problem is precisely eq. (\ref{Rastelli-way}).

\section{Mellin-Barnes representation of Triangle Integrals}\label{Triangle Integral}
In this section, we give a short review on finding Mellin-Barnes representation of one and two-loop triangle integrals following \cite{Usyukina:1992jd, Allendes:2012mr}. 
\subsection*{One-loop Triangle Integral}
\begin{figure}[h]
    \centering
    \begin{tikzpicture}[scale=0.6, every node/.style={scale=0.7}]
    \draw[ultra thick, blue] (0,0) -- (4,0)-- (2,3.464)--(0,0);
    
    \draw[ultra thick, blue] ( -1.732,-1) -- (0,0);
    \draw[ultra thick, blue] (4,0)--(5.732,-1);
     \draw[ultra thick, blue] (2,3.464) -- (2,3.464+2);
     
     \node at (2,-0.5) {\scalebox{1.3}{$\nu_3$}};
     \node at (3.6,1.732) {\scalebox{1.3}{$\nu_2$}};
      \node at (0.5,1.732) {\scalebox{1.3}{$\nu_1$}};

       \node at (2.5,3.464+1) {\scalebox{1.3}{$p_3$}};
       \node at (-1,-0.07) {\scalebox{1.3}{$p_2$}};
       \node at (4.7,-0.85) {\scalebox{1.3}{$p_1$}};
\end{tikzpicture}
    \caption{One-loop triangle diagram}
    \label{fig:one-loop}
\end{figure}
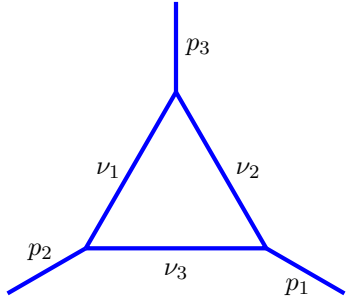
Feynman integral corresponding to one-loop triangle diagram Fig. \ref{fig:one-loop} with  massless internal propagators is
\begin{equation}
    J(d,\nu_1,\nu_2,\nu_3)=\int\frac{d^dk}{((q_1+k)^2)^{\nu_1}((q_2+k)^2)^{\nu_2}((q_3+k)^2)^{\nu_3}}
\end{equation}
where, $q_3-q_2=p_1$,$q_1-q_3=p_2$,$q_2-q_1=p_3$ and all external momenta are considereed to be ingoing ($p_1+p_2+p_3=0$). $\nu_i$ s are introduced for analytic regularization of the diagram which will be important later. For now, clearly, $\nu_i=1$. Finding corresponding Mellin-barnes integral \cite{Usyukina:1992jd} consists of four basic steps of Integrating in and out process :

\vspace{2mm}
\textbf{Step 1: Integrating in Feynman parameters}
\begin{equation}
 \begin{split}
     J(4;1,1,1)&=\int \frac{d^4k}{(q_1+k)^2(q_2+k)^2(q_3+k)^2}\\&=\int d^4k\int dx_1dx_2dx_3\delta(\sum x_i-1)\frac{2!}{[x_1(q_1+k)^2+x_2(q_2+k)^2+x_3(q_3+k)^2]^3}
     \end{split}
 \end{equation}
 Now,
 \begin{equation}
     \begin{split}
    &x_1(q_1+k)^2+x_2(q_2+k)^2+x_3(q_3+k)^2]^3\\&=k^2(x_1+x_2+x_3)+2k(q_1x_1+q_2x_2+q_3x_3)+(x_1q_1^2+x_2q_2^2+x_3q_3^2)\\&=(k+\Vec{q}.\Vec{x})^2+\Delta\quad \text{Taking}\quad x_1+x_2+x_3=1 \quad\text{for the delta function}   
     \end{split}
 \end{equation}
 where $$\Delta  =(x_1q_1^2+x_2q_2^2+x_3q_3^2)-(q_1x_1+q_2x_2+q_3x_3)^2$$
 
\textbf{ Step 2: Integrating out loop-momentum }
 \begin{equation}
     \begin{split}
         &\int \frac{d^4k}{[(k+\Vec{q}.\Vec{x})^2+\Delta]^3}=iS_4 \int \frac{k_E^3dk_E}{(-k_E^2+\Delta)^3}=\frac{iS_4}{2}\int_{0}^{1} \frac{\frac{-\Delta(1-x)}{x}\times \frac{\Delta}{x^2}dx}{(\frac{\Delta}{x})^3}=-\frac{i\pi^2}{2\Delta}
     \end{split}
 \end{equation}
 Here $S_4$ is volume of four-sphere. Now
 \begin{equation}
 \begin{split}
 \Delta&=(x_1q_1^2+x_2q_2^2+x_3q_3^2)-(q_1x_1+q_2x_2+q_3x_3)^2 \\&=[x_1x_2p_3^2+x_1x_3p_2^2+x_2x_3p_1^2] \quad \text{[Using $x_1+x_2+x_3=1$]}
 \end{split}
 \end{equation}
 
 Hence,
\begin{equation}
    J(4;1,1,1)=-i\pi^2\int_{0}^{1}\int_{0}^{1}\int_{0}^{1}\frac{dx_1dx_2dx_3\delta(\sum x_i-1)}{[x_1x_2p_3^2+x_1x_3p_2^2+x_2x_3p_1^2]}
 \end{equation}

\vspace{2mm}
\textbf{Step 3 and 4: Integrating in Mellin variables and Integrating out Feynman parameters}\\

We can now use Mellin-Barnes integral to rewritwe as:
\begin{equation}
    \begin{split}
        &J(4;1,1,1)\\&=-i\pi^2\int_{0}^{1}\int_{0}^{1}\int_{0}^{1} dx_1dx_2dx_3 \delta(\sum x_i-1) \frac{1}{x_1x_2p_3^2} \int \frac{dsdt}{(2\pi i)^2} (\frac{x_2x_3p_1^2}{x_1x_2p_3^2})^s(\frac{x_1x_3p_2^2}{x_1x_2p_3^2})^t\Gamma(-s)\Gamma(-t)\Gamma(1+s+t)\\&=-\frac{i\pi^2}{p_3^2}\int \frac{dsdt}{(2\pi i)^2} x^s y^t \Gamma(-s)\Gamma(-t)\Gamma(1+s+t)  \int_{0}^{1}\int_{0}^{1}\int_{0}^{1} dx_1dx_2dx_3 \delta(\sum x_i-1)x_1^{-s-1}x_2^{-t-1}x_3^{s+t}\\&=-\frac{i\pi^2}{p_3^2}\int \frac{dsdt}{(2\pi i)^2} x^s y^t \Gamma^2(-s)\Gamma^2(-t)\Gamma^2(1+s+t)
    \end{split}
\end{equation}
where, $$x=\frac{p_1^2}{p_3^2}\quad\text{and}\quad y=\frac{p_2^2}{p_3^2}$$
Hence, 

\begin{equation}
\boxed{  J(4;1,1,1)=-\frac{i\pi^2}{p_3^2}\int \frac{dsdt}{(2\pi i)^2}\, x^s y^t \,\Gamma^2(-s)\Gamma^2(-t)\Gamma^2(1+s+t)} 
\end{equation}

\subsection*{Two-loop Triangle Integral}

\begin{figure}[h]
    \centering
    \begin{tikzpicture}[scale=0.6, every node/.style={scale=0.7}]
    \draw[ultra thick, blue] (0,0) -- (3.464,2) -- (3.464,-2) -- (0,0);
    \draw[ultra thick, blue] (0,0) -- (-2,0);
    \draw[ultra thick, blue] (3.464,2) -- (3.464+1,2+1.732);
    \draw[ultra thick, blue] (3.464,-2) -- (3.464+1,-2-1.732);
     \draw[ultra thick, blue] (1.732,1) -- (1.732,-1);

      \node at (-1,-0.45) {\scalebox{1.3}{$p_3$}};
      \node at (3.464+1,-2.8) {\scalebox{1.3}{$p_2$}};
      \node at (3.464+0.1,+2.9) {\scalebox{1.3}{$p_1$}};

      \node at (4.3,0) {\scalebox{1}{$(1+\epsilon_3)$}};
        \node at (2.6,0) {\scalebox{1}{$(1+\epsilon_3)$}};
        
        \node at (0.6,1) {\scalebox{1}{$(1+\epsilon_1)$}};
        \node at (0.6,-1) {\scalebox{1}{$(1+\epsilon_2)$}};

        \node at (0.6+1.6,1+1) {\scalebox{1}{$(1+\epsilon_2)$}};
         \node at (0.6+1.6,-1-1) {\scalebox{1}{$(1+\epsilon_1)$}};
    \end{tikzpicture}
    \caption{Two-loop triangle diagram}
    \label{fig:two-loop}
\end{figure}
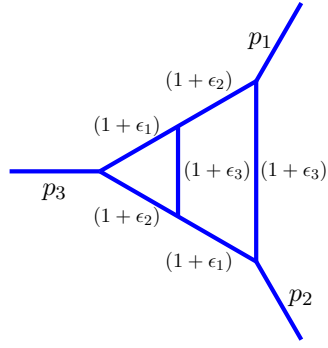

We will denote the notation $$\int_2(\epsilon_1,\epsilon_2,\epsilon_3)$$ to the 2-loop momentum integral with incoming momenta $p_1, p_2, p_3$ as shown in the figure \ref{fig:two-loop}. Note that we have used a special analytical regularization scheme, where we replace unit powers of denominators by $(1+\epsilon_i)$, provided that $\epsilon_1+\epsilon_2+\epsilon_3=0$.

The key formula in this calculation is Eq. (25) in \cite{Usyukina:1992jd} :
\begin{equation}\label{epsilon eq}
    \begin{split}
        &\int_2(\epsilon_1,\epsilon_2,\epsilon_3)=\frac{J}{(p_3^2)^{1-\epsilon_3}}[\frac{1}{\epsilon_2\epsilon_3}(p_2^2)^{\epsilon_3}J(1,1,1-\epsilon_1)+\frac{1}{\epsilon_1\epsilon_2}J(1,1,1+\epsilon_3)+\frac{1}{\epsilon_1\epsilon_3}(p_1^2)^{\epsilon_1}J(1,1,1-\epsilon_2)]\\ &J=\frac{\Gamma(1-\epsilon_1)\Gamma(1-\epsilon_2)\Gamma(1-\epsilon_3)}{\Gamma(1+\epsilon_1)\Gamma(1+\epsilon_2)\Gamma(1+\epsilon_3)}
    \end{split}
\end{equation}
This relation has been rederived recently in \cite{Allendes:2012mr}. It connects the two-loop result with the one-loop one. Since mellin transform of $J(\nu_1,\nu_2,\nu_3)(\equiv J(d;\nu_1,\nu_2,\nu_3))$ is known \cite{Ussykina-1},(we calculated it for the simple case $\nu_i=1$ in the previous section)
\begin{equation}
    J(\nu_1,\nu_2,\nu_3)=\frac{i\pi^2}{(p_3^2)^{\sum \nu_i-d/2}}\int dz_2dz_3 x^{z_2}y^{z_3}D^{(z_2,z_3)}[\nu_1,\nu_2,\nu_3]
\end{equation}
where x and y are the same as defined in the last section and 
\begin{equation}
\begin{split}
    D^{(z_2,z_3)}[\nu_1,\nu_2,\nu_3]=&\frac{\Gamma(-z_2)\Gamma(-z_3)\Gamma(-z_2-\nu_2-\nu_3+d/2)\Gamma(-z_3-\nu_1-\nu_3+d/2)}{\prod_{i}\Gamma(\nu_i)}\\&\times \frac{\Gamma(z_2+z_3+\nu_3)\Gamma(\sum \nu_i-d/2+z_3+z_2)}{\Gamma(d-\sum_i\nu_i)}
    \end{split}
\end{equation}
Clearly from the above Eq. (\ref{epsilon eq}), we can get Mellin transform of two-loop triangle. The result is found in \cite{Allendes:2012mr}:
\begin{equation}
    \left[\frac{D^{(u,v-\epsilon_2)}[1-\epsilon_1]}{\epsilon_2\epsilon_3}+\frac{D^{(u,v)}[1+\epsilon_3]}{\epsilon_1\epsilon_2}+\frac{D^{(u-\epsilon_1,v)}[1-\epsilon_2]}{\epsilon_1\epsilon_3}\right]
\end{equation}
Hence, two loop triangle diagram: 
\begin{equation}
    \begin{split}
        C^{(2)}(p_1^2,p_2^2,p_3^2)=(\frac{i\pi^2}{p_3^2})^2\int ds dt \, x^s y^t M^{(2)}(s,t)
    \end{split}
\end{equation}
where, \begin{equation}
\begin{split}
    M^{(2)}(s,t)&=\lim_{\epsilon_2 \to 0, \epsilon_1 \to 0}J[\frac{D^{(s,t-\epsilon_2)}[1-\epsilon_1]}{\epsilon_2\epsilon_3}+\frac{D^{(s,t)}[1+\epsilon_3]}{\epsilon_1\epsilon_2}+\frac{D^{(s-\epsilon_1,t)}[1-\epsilon_2]}{\epsilon_1\epsilon_3}]\\&=
    \frac{1}{2}\Gamma^2(-s) \Gamma^2(-t) \Gamma^2(1+s+t)[(\Psi'(-t)+\Psi'(-s))-(\Psi(-t)-\Psi(-s))^2-\pi^2]\label{eq:M}
    \end{split}
\end{equation}
with $\Psi(z)$ being digamma function of $z$.



\end{document}